\documentclass{ifacconf}

\usepackage{graphicx, indentfirst}
\usepackage{natbib}        
\usepackage{amsmath}
\usepackage{amssymb}

\begin{document}

\begin{frontmatter}

\title{Detection of screw threads in computed tomography 3D density fields\thanksref{footnoteinfo}} 

\thanks[footnoteinfo]{Presented images were rendered using Linderdaum Engine, http://www.linderdaum.com.}

\author[First]{Sergey Kosarevsky} 
\author[Second]{Viktor Latypov} 

\address[First]{Saint-Petersburg Institute of Mechanical-Engineering, Saint-Petersburg, Russia 195197 (e-mail: kosarevsky@mail.ru).}
\address[Second]{Saint-Petersburg State University, Saint-Petersburg, Russia 198504.}

\begin{abstract}                
In this paper, a new method is proposed to automatically detect screw threads in 3D
density fields obtained from computed tomography measurement devices. The described 
method can be used to automate many operations during screw thread inspection process and 
drastically reduce operator's influence on the measurement process resulting in lower 
measurement times and increased repeatability.
\end{abstract}


\end{frontmatter}

\maketitle

\section{Introduction}
\indent

\indent
Many types of coordinate measurement systems are widely used to inspect parameters
of metric screw threads~\citep{NPLNotes}.
Screw thread measurement using coordinate measuring machines (CMM) is a tedious process
which requires high operator's skill. CMM measurements usually aquire a limited set of points
suitable for a feature-based inspection, however not enough for an in-depth analysis and
assesment of a screw pair fit. This limitation is crucial while dealing with screw threads
for mission critical applications (petrochemical industry, heavy engineering industry et al.).
Nowdays, as a result of well-established international practice, the complex inspection of
screw threads is performed using thread gauges. Gauges are subject to wear and tear and
require regular replacement and inspection which yields expenses.

\begin{figure*}[!ht]
   \begin{center}
      \includegraphics[width=10cm]{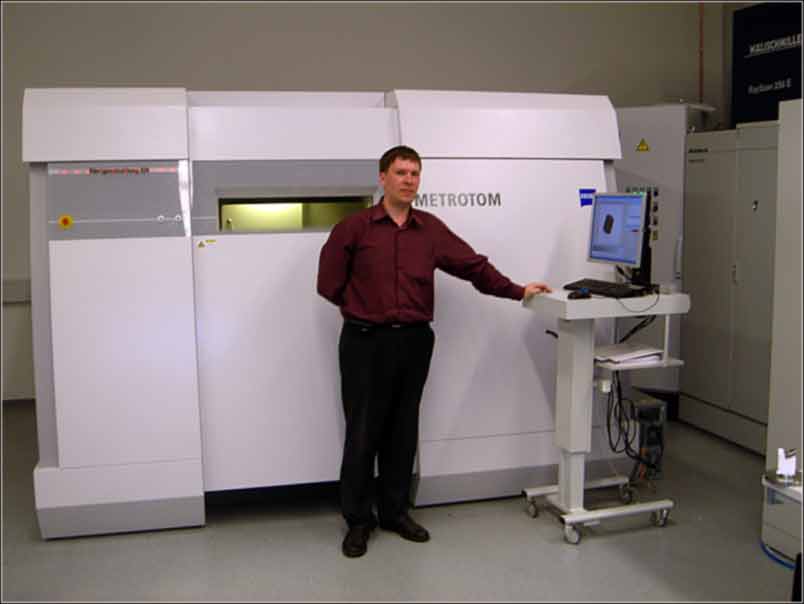}
   \end{center}
   \caption{X-ray CT scanner Carl Zeiss Metrotom 1500}
   \label{Metrotom}
\end{figure*}

\indent
In this paper, a new method is proposed, allowing to perform
feature-based inspection and complex in-depth analysis of screw
threads using the date aquired from computed tomography (CT) measuring devices.
Experimental part of this work was done using X-ray CT scanner Metrotom~1500~(see Fig.\ref{Metrotom})
by Carl Zeiss~IMT~GmbH.

\section{Related work}

\indent
Threaded and plain gauges are widely used for screw thread inspection during recent decades. They
allow complex inspection of threaded fits~\citep{NPLNotes}. Gauges are expensive inspection instrument
since they are subject to wear and the whole inspection procedure can be long-lasting
for large diameter threads (M150 and above). Thread gauges that are proved to be out-of-range by direct measurements
can be fit with good master gauges. This situation is typical but gauges and direct measurements are not
mutually exclusive methods. Considering the NPL experience~\citep{NPLNotes}
one can use both master gauges and direct measurements to do an elementwise
inspection of the thread gauges.
First, one should measure pitch, pitch diameter and inner/outer diameters of the thread.
Thereupon gauges should be applied to assist the direct measurements. Elementwise inspection is more
accurate and can yield numerical results (instead of simple go/no-go answer).
The point is direct elementwise measurements do not evaluate the whole surface of the thread therefore
form deviations can be overlooked. Gauges can prevent this kind of errors even after elementwise measurements
succeeded.

Gauges are unable to separate pitch and diameter deviations. In addition, pitch errors
can be hidden by increased pitch diameter. It is traditionally considered practical to use
gauges for small threads. As described in NPL notes~\citep{NPLNotes} the primary reason
for this discrepancy between gauges and direct measurements is thread form deviations.

\indent
Nowadays, a lot of elementwise thread assessment methods have evolved. Their majority is
based on coordinate measurement devices and numeric evaluation of the results. A lot of
researchers work torwards the improvement of thread inspection. In~\citep{Ikon95}
the methos of position and form measurement is proposed based virtual gauges.
Carmignato and Chiffre~\citep{Carm03} proposed a screw inspection method with 
a special needle-like probe fitted on a coordinate measuring machine.
In~\citep{He06} authors perform optical inspection of damaged screw threads
using CCD camera.
Many techniques usually involve different shape analysis algorithms 
from the field of computer vision and pattern recognition.
Robertson and Fisher~\citep{Robert01} experimented with 3D scanners 
and their application to large thread measurement.
They examined that it is possible to extract parameters of screw
threads from 3D scanned data. However they deal only with inner
and outer radii of the thread.
In the proceedings~\citep{Katz09} a method is presented
to perform an elementwise inspection of internal threads using laser sensor
and CCD camera.
Kosarevsky and Latypov~\citep{Kos10-11} used Hough transformation
to extract features from planar sections obtained via profile measuring machines.
The mathematics behind these image recognition techniques can be found in~\citep{Prin92}. \\

\indent
In~\citep{Perng10} a working system is proposed capable of automatic internal thread inspection.
It is based on industrial endoscope and computer vision
algorithms. However, the main purpose of this system is to detect surface defects and not its geometrical
properties. An in-depth overview of classical thread inspection methods is provided in NPL Notes on Screw Gauges~\citep{NPLNotes}.
Current element-wise thread inspection methods are mainly based on the work of G.Berndt~\citep{Berndt40}. \\

\indent
Application of high resolution computer tomography (CT) in dimensional metrology has grown popular during the recent years.
It has moved from qualitative assessment of workpieces in non-distructive tests to a precise measurement instrument~\citep{DeChiffre2011}.

\indent
The accuracy of spatial measurements in these tasks is highly dependent on the geometry and material workpieces.
A lot of numeric compensation algorithms are used to achieve the highest possible accuracy and reduce
different artifacts of CT scanning.

\indent
Traceability of coordinate measurements obtained from computed tomography devices
is ensured via special calibration gadgets and procedures. Nowadays these methods
received ISO certification. Modern software can extract geometrical features from
measured density fields and evaluate their parameters according to ISO norms.
To assess the accuracy of CT systems (besides calibration) sets of reference workpieces
are used. These workpieces are calibrated on high-precision CMMs and results are compared
to the CT data. Recent experiments on GE Sensing \& Inspection Technologies (Germany) CT devices show~\citep{Luebbehuesen09}
the deviation of results for distances and diameters to be within 6~$\mu m$. Carl Zeiss Metrotom 800 X-ray CT scanners
can achieve values of ~$ MPE_{E} = 4.5 + 0.01 L $~$\mu m$~\citep{Benniger09}.

\section{Initial approach}

\indent
In this paper, a method of planar sections is used to numericaly assess the quality of screw threads. It allows
to reduce the problem of 3D shapes recognition to the finite number of 2D recognition problems.
Our method operates on a bundle of planes that contain thread axis. 
The principal symmetry axis of the CT-scan is considered to be the thread axis.
Performing image recognition in each plane a numeric evaluation of thread parameters can be obrained.
The planar evaluation algorithm was presented in~\citep{Kos10-11}. Using the obtained results one can assess 
the quality of the measured thread.

\indent
Let $f(x_1,x_2,x_3): \mathbb{R}^3 \mapsto \mathbb{R}$ be the material density distribution inside the object measured by a CT device and let
$ \mathcal{D} \subset\mathbb{R}^3$ be the domain where function $f$ is defined.
Let us consider a bundle of planes $P_\alpha$ that share a common point $r=(\bar{x}_1,\bar{x}_2,\bar{x}_3)$
(``mass center'' of the measured object) as ``distinguished'' planar sections of the thread. Components $r$ are evaluated as follows
\begin{equation}
\bar{x}_i = \frac{\int_{\mathcal{D}} x_i f(x_1,x_2,x_3)\; dx_1\; dx_2\; dx_3}{\int_{\mathcal{D}} f(x_1,x_2,x_3)\; dx_1\; dx_2\; dx_3},\mbox{  }i=0,1,2.
\end{equation}

\indent
The thread axis is the common axis of planes $P_{\alpha}$ which is the principal symmetry axis of the scanned object.
Symmetry axes are approximated with the eigenvectors of the covariance matrix
\begin{equation}
\label{covariance}
A=\left(\begin{array}{ccc}
\mu_{200} & \mu_{110} & \mu_{101} \\
\mu_{110} & \mu_{020} & \mu_{011} \\
\mu_{101} & \mu_{011} & \mu_{002}
\end{array}
\right)
\end{equation}
where
\begin{center}
\begin{equation*}
\mu_{ijk}=\int_{\mathcal{D}}W_{ijk} f(x_1,x_2,x_3)\;dx_1\;dx_2\;dx_3,
\end{equation*}
\begin{equation}
\label{Momentum}
W_{ijk}=\left(x_1-\bar{x}_1\right)^i \left(x_2-\bar{x}_2\right)^j \left(x_3-\bar{x}_3\right)^k.
\end{equation}
\end{center}

\begin{figure*}[!ht]
   \begin{center}
      \includegraphics[width=10cm]{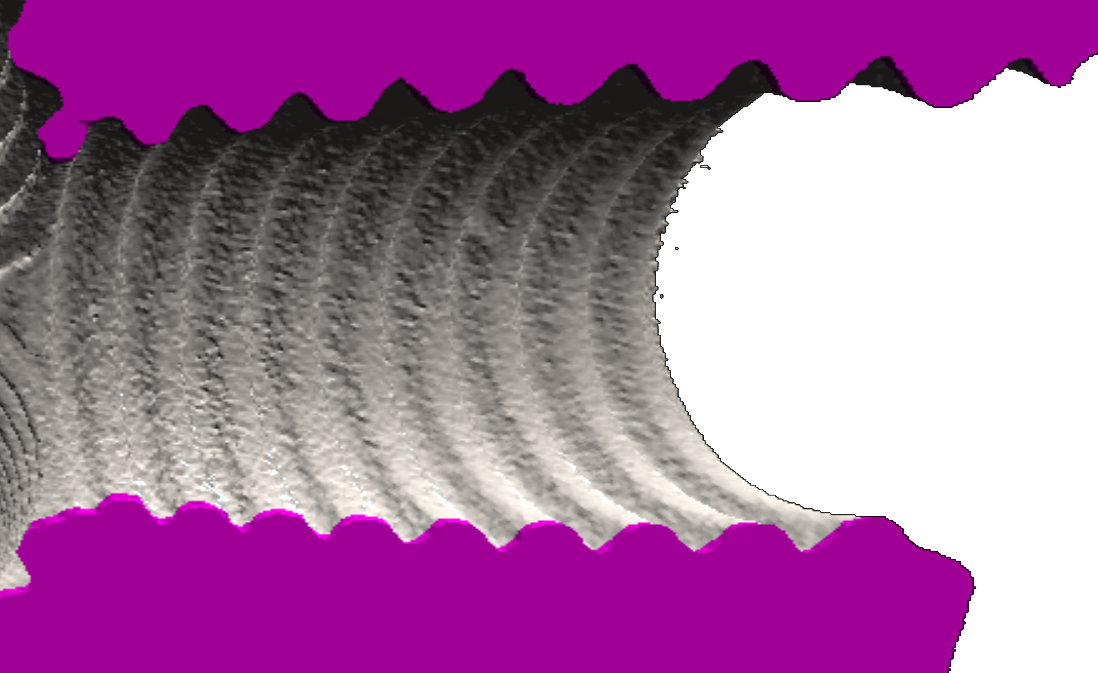}
   \end{center}
   \caption{Planar section passing through the axis of M5 thread}
   \label{LongSection}
\end{figure*}

\indent
The plane $P_{\alpha}$ passing through the axis of the internal M5 thread is shown in Fig.~\ref{LongSection} with the section of the object.

\section{Analytic representation of a screw thread}

\indent
In~\citep{Nicolson91} and~\citep{Nicolson93} Nicolson et al. proposed a model to represent a basic profile of the metric screw surface.
Their model is based on piecewise functions and is effective at finding contact surfaces.
In this work a method more suitable for thread recognition is used.

\indent
Consider the family of helicoidal surfaces

\begin{equation}
\label{CoordSurfaces}
\left\{
\begin{array}{lcl}
  x&=& D^{-1} w \left( D + \left| u \right| H \right) \sin{(R v + \pi u )}   \\
  y&=& D^{-1} w \left( D + \left| u \right| H \right) \cos{(R v + \pi u )}   \\
  z&=& v
\end{array}
\right.
\end{equation}
paramerized by $u \in [-1;1], v \in [0; V]\;\; \mbox{and}\;\; w \in [0; +\infty)$.

\indent
Assumption $ w = D $ yields the surface in Fig.\ref{Helicoid}:

\begin{figure*}[!ht]
   \begin{center}
      \includegraphics[width=15cm]{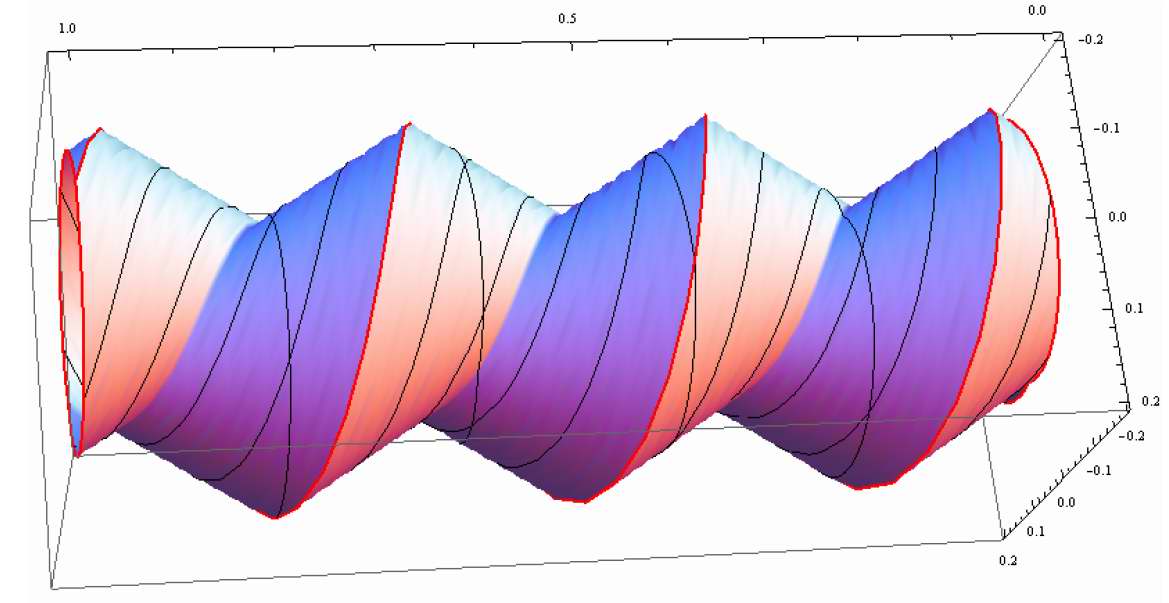}
   \end{center}
   \caption{Surface $ w = D $}
   \label{Helicoid}
\end{figure*}

\indent
The transform $ (x; y; z) \to (u; v; w) $ can be used to check if some point $ (x; y; z) $ belongs to the surface $ w $ = $ D $.
From (\ref{CoordSurfaces}) it follows that $ v = z $. Also,
$ x^{2} + y^{2} = D^{-1} w^{2} \left( D + |u| H \right)^{2} $ and
$ x/y = \tan{(R v + \pi u)} $, which gives

\begin{center}
\begin{equation}
\begin{array}{lcl}
   u &=& \pi^{-1}\left(\arctan{\left(y^{-1}x\right)}-R z\right) \\
   v &=& z \\
   w &=& D\left(D+|u|H\right)^{-1}\sqrt{x^{2}+y^{2}}.
\end{array}
\end{equation}
\end{center}

\indent
It is easy to see that planar sections of the surface $ w = D $ are piecewise linear.
Projection on $ Oyz $ plane yields $ x = 0 $, that is $ \sin{(R v + \pi u)} = 0 $ or

\begin{equation}
\left\{
\begin{array}{lcl}
   R v + \pi u = \pi k, u = \pi^{-1}\left(\pi k - R v\right), \\\\
   y = (-1)^k w D^{-1} \left( D + |u| H \right), \\\\
   z = v,
\end{array}
\right.
\end{equation}
and $u$ is selected from the $[-1;1]$ segment by the appropriate choice of the $k$ value.

\begin{figure*}[!ht]
   \begin{center}
      \includegraphics[width=12cm]{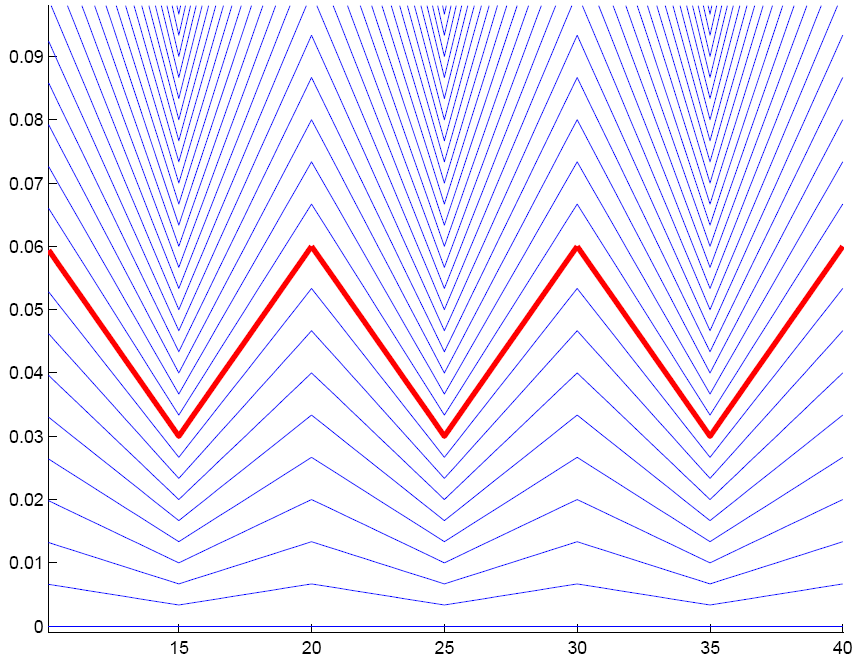}
   \end{center}
   \caption{Section of the surface $ w = D $ at $ x = 0 $}
   \label{Sections}
\end{figure*}

\indent
A bundle of coordinate surfaces intersecting the plane $ x = 0 $ is shown in Fig.\ref{Sections}.

\begin{figure*}[!ht]
   \begin{center}
      \includegraphics[width=15cm]{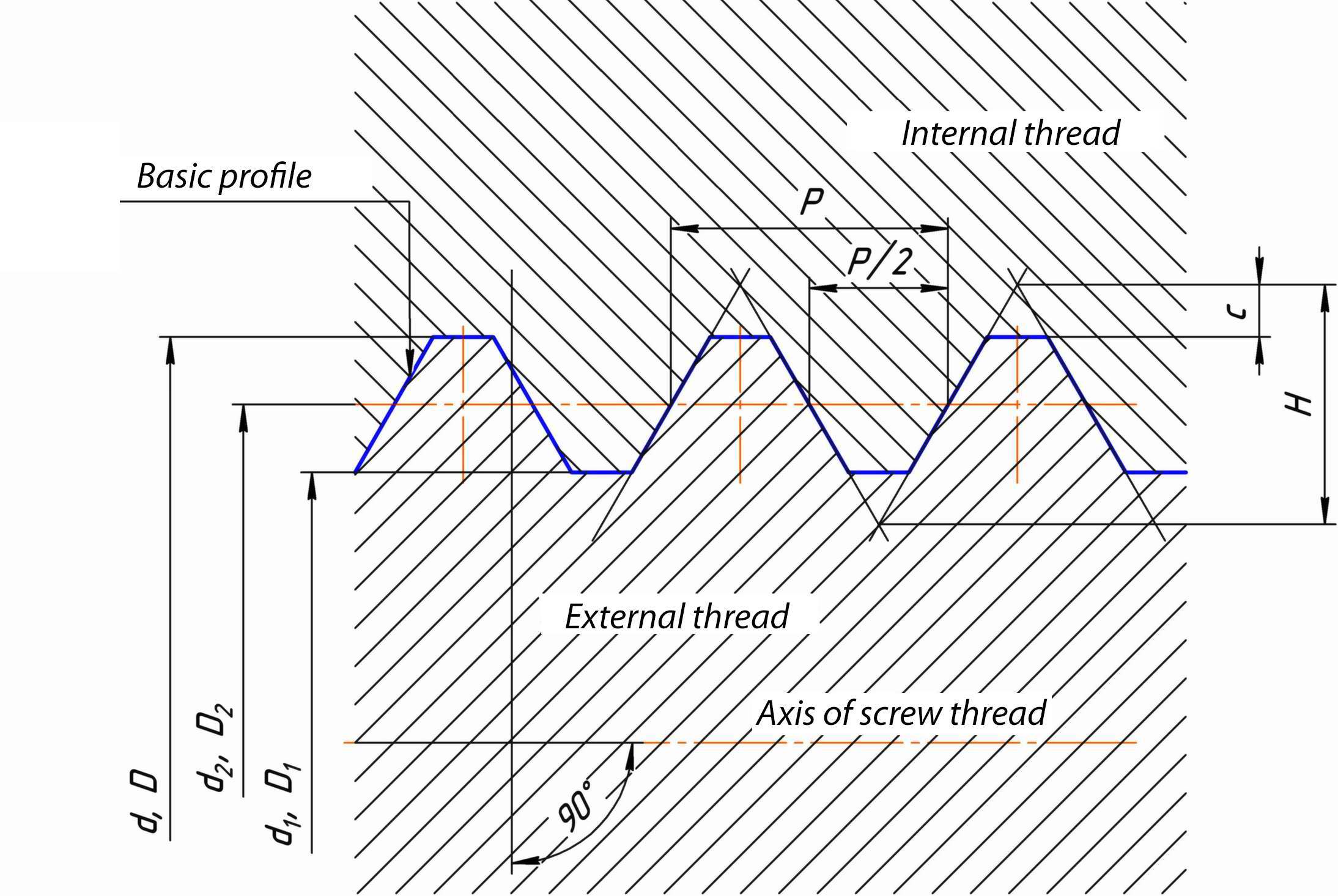}
   \end{center}
   \caption{Basic profile of ISO metric screw thread~\citep{ISO68-1}}
   \label{MetricThread}
\end{figure*}

\indent
Basic profile~\citep{ISO68-1} of the metric screw thread (Fig.\ref{MetricThread}) is defined by the pitch $P$, the height of generating triangle $H$
and the nominal diameter $D$.

\indent
According to~\citep{ISO68-1} the values are:

\begin{equation}
   c = \frac{1}{8} H  \hspace{0.7cm}  D_{1} = D - 2\cdot \frac{5}{8} H  \hspace{0.7cm}  D_{2} = D - 2\cdot \frac{3}{8} H
   \label{ThreadEq1}
\end{equation}

\indent
For a metric screw thread the profile angle is $ \alpha = 60^{o} $, so that
$ \tan{\frac{\alpha}{2}} = \frac{P}{2 H} $ and equations (\ref{ThreadEq1}) yield:

\begin{equation*}
   c = \frac{P}{16 \tan{\frac{\alpha}{2}} }  
\end{equation*}

\begin{equation*}
   D_{1} = D - \frac{5}{8} \frac{P}{\tan{\frac{\alpha}{2}} }
\end{equation*}

\begin{equation*}
   D_{2} = D - \frac{3}{8} \frac{P}{ \tan{\frac{\alpha}{2}} }
\end{equation*}

\indent
Any longitudinal planar section of the nominal screw thread is the planar section of the surface Fig.~\ref{Sections},
truncated on both sides according to the parameter $ c $.

\section{Assesment of screw thread parameters}

\begin{figure*}[!ht]
   \begin{center}
      \includegraphics[width=10cm]{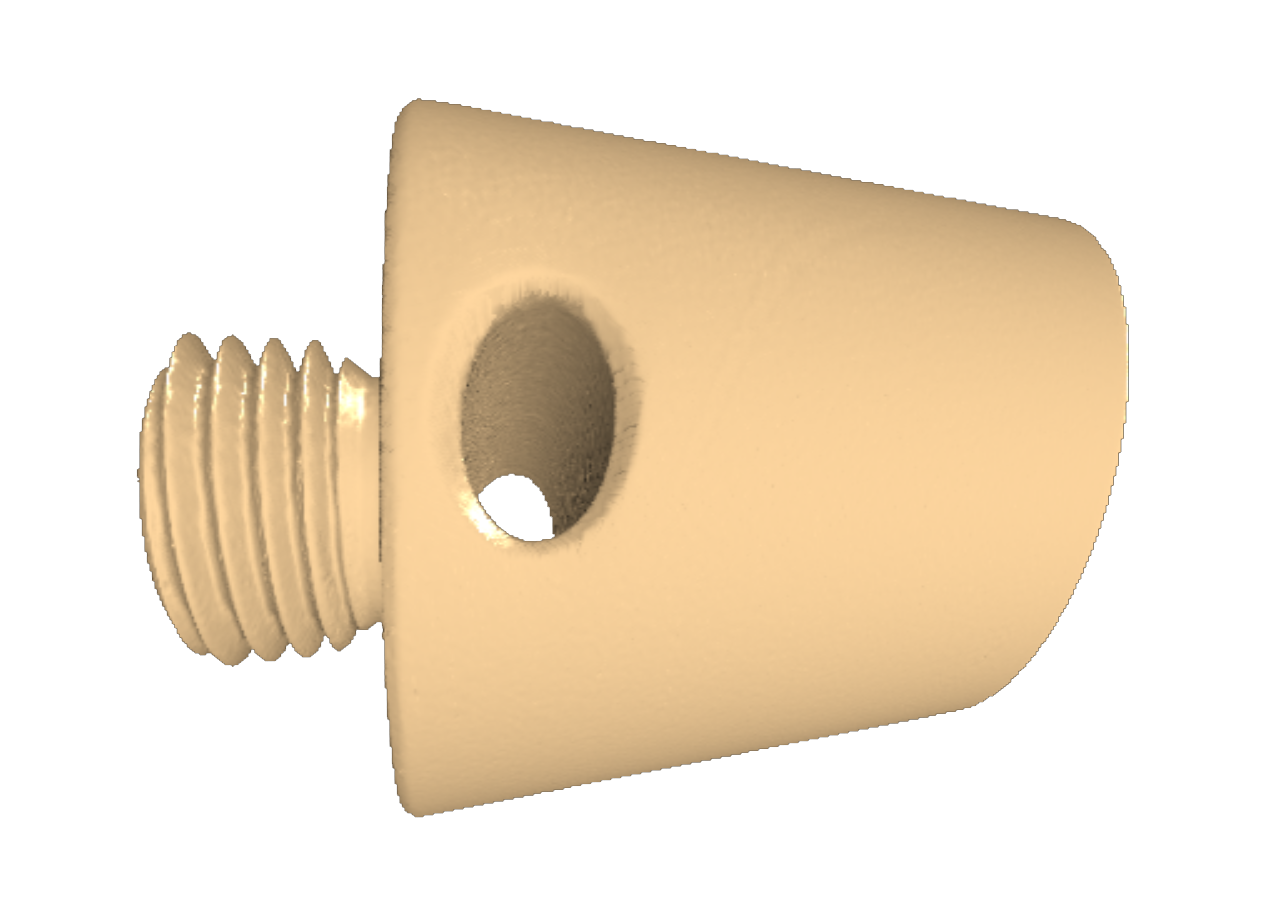}
   \end{center}
   \caption{M5 screw thread density field after the thresholding filter application}
   \label{Segmentation}
\end{figure*}

\begin{figure*}[!ht]
   \begin{center}
      \includegraphics[width=10cm]{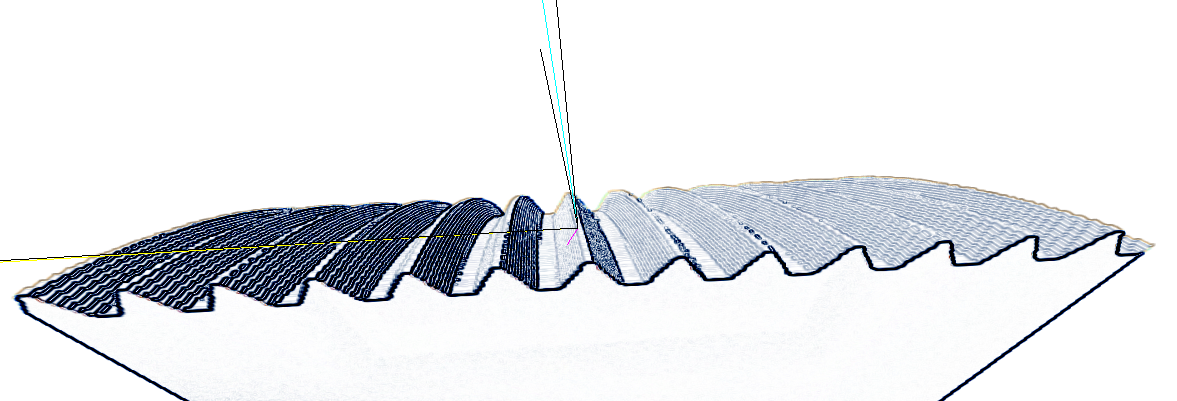}
   \end{center}
   \caption{Results of 3D Sobel operator application to the region of density field}
   \label{3DSobel}
\end{figure*}

\indent
The overall workflow of our algorithm is as follows:

\begin{enumerate}
   \item {\bf Acquire scalar 3D density field.} Authors used Carl Zeiss Metrotom 1500 X-Ray CT scanner to acquire data.
			Also, several sets of generated data were used to perform an in-depth data analyzis and proove the correctness
			of our implementation.

   \item {\bf Perform feature extraction.} Apply the thresholding filter. Non-zero values correspond to in-material points.
         Fig.~\ref{Segmentation} shows the thresholding result for the scanned external M5 thread with a threshold value of 0.48.
         The actual range of interest is specified manually.

   \item {\bf Establish thread axis.} Calculate covariance matrix using the equation~(\ref{covariance}) and determine its eigenvectors. One of the eigen vectors defines
			as approximate thread axis, which is a common line for all of the 2D cross section planes.

   \item {\bf Extract isosufrace.} Apply the three-dimensional Sobel filter to determine the points of thread surface. The result of 3D Sobel operator is shown in Fig.~\ref{3DSobel}.

   \item {\bf Generate planar sections.} Fix an arbitrary plane containing the center of mass of the volume and previously 
         determined thread axis. Rotate this plane using a number of uniformly distributed angles in the $0\ldots 2\pi$ range.
         Clusterize points into buckets corresponsing to the closest plane. Every single point can get into a bucket only once.

   \item {\bf Process each planar section (bucket) as described in~\citep{Kos10-11}.} Straight lines representing flanks
         of the thread are obtained via Hough transformation. Sort obtained lines and find intersection points. Intersection points
         closest to the profile average line are used to fit circles into screw thread groves. Fit circles into the groves. Radii of the circles 
			are chosen so that circles contact the profile close to the pitch diameter, as in the three-wires method.

   \item {\bf Evaluate thread pitch and pitch diameter.} Distances between the 
         fitted circles are used for assessment of screw thread parameters in every planar section.
\end{enumerate}

\indent
The described procedure can be applied not only to density fields, but also to
3D point clouds. In that case segmentation is already done and one can continue
from step 3, using the coordinates of separate points.

\section{Experiments and autogenerated models}

\indent
The correctness of the algorithm is checked by processing a number of
generated volumes corresponding to some screw threads with specified parameters.
The generated volume data is an $1024^3$ 8-bit three-dimensional array of samples
which is essentialy a quantized graph of the $\chi_{S}$ function corresponding to the screw thread.
To estimate the precision of the algorithm the generated volume data is corrupted
with gaussian noise.

\begin{figure*}[!ht]
   \begin{center}
      \includegraphics[width=10cm]{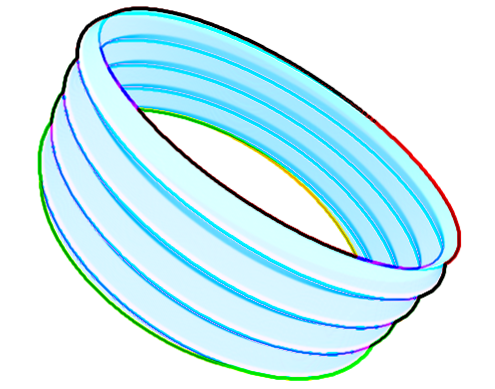}
   \end{center}
   \caption{Autogenerated screw thread model}
   \label{Autogenerated}
\end{figure*}

The sample of the generated screw thread model is presented in Fig.~\ref{Autogenerated}.

\indent
Two aluminum samples with metric screw threads were scanned using Carl Zeiss Metrotom 1500 CT X-ray scanner: internal and external M5 ISO screw threads.
Numeric results of screw thread pitch evaluation using 36 planar sections are presented in the table~\ref{NumericResults}.

\begin{table}
\begin{center}
\caption{Numeric results}
\label{NumericResults}
\begin{tabular}{cp{2cm}p{2cm}}
\hline
Thread  & Pitch (min.max.), mm & Pitch std. dev., mm \\
\hline
External thread M5   & $ 0.97 \div 1.01 $ & $ 0.02 $  \\
Internal thread M5   & $ 0.91 \div 1.07 $ & $ 0.06 $  \\
Artificial model  & $ 1.00 \div 1.00 $ & $\approx 0.00$ \\
\hline
\end{tabular}
\end{center}
\end{table}

\indent
Significant value of standard deviation for internal thread is due to thicker material and size of the workpiece compared to the
external thread workpiece. Standard deviation for our artificial model was below~0.001~mm. However, probability distribution
and dependence on the input data need to be assessed more carefully.

\indent
Our software was implemented in C++ using Minpack~\citep{Minpack} optimization library and KHT Sandbox (https://sourceforge.net/projects/khtsandbox)~\citep{Fern08},
the reference implementation of the kernel-based Hough transform for detecting straight lines in binary images.
It allows a software implementation to achieve real-time performance even on relatively large images.
Fast evaluation of $f(x_1,x_2,x_3)$ gradients was done in a GLSL shader.

\indent
Direct evaluation of moments $\mu_{ijk}$ using the formula~(\ref{Momentum}) (necessary to establish thead axis),
can take significant time at higher resolution. However, in this paper, the problem is not addressed. In case where material density
corresponds to a metric screw thead\footnote{Segmented volume contains only binary 0 and 1 without any intermediate values.}
special optimizations methods can be used~\citep{FlusserSuk93}.

\indent
Eigenvectors of a covariance matrix are calculated using Jacobi eigenvalue algorithm~\citep{Flannery92}.

\indent
Presented images were rendered using Linderdaum Engine and Linderdaum Volume Rendering Library.

\section{Future work}

\indent
Future work must be undertaken to assess the measurement uncertainty
of the proposed algorithm and how noises in the input data and their filtering
affect the results. Measurement uncertainty can be determined via Monte-Carlo trials according
to~\citep{GUM} and~\citep{GUM-Suppl}. \\

\indent
It is important to mention that finding values of parameters $P$ and $D$ so that
the virtual gauge will fit into the measured thread without intersection will
enable direct evaluation of virtual pitch diameter of the thread.

\section{Conclusions}

\indent
One of the main advantages of the proposed method is that the result
is not just a binary ``go/not-go``, but contains additional numeric data
which can be evaluated to determine the exact thread features with deviations.
It is important for further technological decisions concerning the production of threaded parts.

\indent
One of the main flaws of the proposed methods is the scanning speed
of CT equipment. It is required up to one hour of machine time to inspect a single
workpiece. However, other types of surface scanning devices (i.e. laser scanning of external threads)
can yield improved performance and provide usable 3D point clouds.

\section{Acknowledgements}

\indent
Authors would like to acknowledge the funding assistance of the OPTEC company (representative of
Carl Zeiss in Russia) and would like to thank Wolfgang Schwarz from Carl Zeiss~IMT and 
Peter Hoyer from Carl Zeiss 3D Metrology Services for provided machine time with a Metrotom 1500 CT-scanner
and valuable technical assistance.

\indent
This work was supported by the grants OPTEC --- Carl Zeiss 2010 and OPTEC 2012.
Presented images were rendered using Linderdaum Engine and Linderdaum Volume Rendering Library.

\bibliography{Disser}
                                                                                                                                              
\end{document}